\documentclass[12pt]{article}

\usepackage{pdfpages}

\usepackage{scicite}
\usepackage{times}

\usepackage{amsmath}
\usepackage{amsfonts}
 
\usepackage{booktabs}%%%will fix the table (but maybe we don't need it)
\usepackage{dcolumn}%%%will fix the table (but maybe we don't need it)

\usepackage{enumitem}

\usepackage{rotating}%for landscape figures
\usepackage{graphicx}

\usepackage[para,online,flushleft]{threeparttable}

\newenvironment{sciabstract}{ 
	\begin{quote} \bf }
	{\end{quote}}

\newcounter{lastnote}

 \usepackage{color}
 \definecolor{Blue}{rgb}{0,0,1}
 
 \definecolor{Orange}{rgb}{1,0.5,0}
 
 \definecolor{Green}{rgb}{0,1,0}

 \usepackage[position=t,singlelinecheck=off]{subfig}
 \usepackage{caption}

\usepackage{float}

\title{\vspace{-13mm}Mapping the Privacy-Utility Tradeoff in Mobile Phone Data for Development}

\author
{
	Alejandro Noriega-Campero$^{1\ast}$, 
    Alex Rutherford$^{1}$, 
    Oren Lederman$^{1}$,\\ 
    Yves A. de Montjoye$^{2}$, 
    Alex Pentland$^{1}$ \\ \\
	\normalsize{$^{1}$Massachusetts Institute of Technology, Cambridge, MA, USA}\\
	\normalsize{$^{2}$Imperial College London, London, UK}\\
	\normalsize{$^\ast$ To whom correspondence should be addressed: noriega@mit.edu}
	\vspace{-1.5mm}
}

\date{}

\usepackage[british]{babel}
\usepackage{enumitem}

\begin{document} 
	
	
	\baselineskip18pt    %24pt
	
	
	\maketitle

\begin{sciabstract}    % abstract
	\label{abstract}

Today's age of data holds high potential to enhance the way we pursue and monitor progress in the fields of development and humanitarian action. We study the relation between data utility and privacy risk in large-scale behavioral data, focusing on mobile phone metadata as paradigmatic domain. To measure utility, we survey experts about the value of mobile phone metadata at various spatial and temporal granularity levels. To measure privacy, we propose a formal and intuitive measure of reidentification risk\iffalse in high-dimensional data\fi\textemdash the \textit{information ratio}\textemdash and compute it at each granularity level. 
Our results confirm the existence of a stark tradeoff between data utility and reidentifiability, where the most valuable datasets are also most prone to reidentification. When data is specified at ZIP-code and hourly levels, outside knowledge of only 7\% of a person's data suffices for reidentification and retrieval of the remaining 93\%. In contrast, in the least valuable dataset, specified at municipality and daily levels, reidentification requires on average outside knowledge of 51\%, or 31 data points, of a person's data to retrieve the remaining 49\%. \iffalse However, we also find that the tradeoff is not strict: we identify Pareto-suboptimal granularity levels that substantially increase reidentifiability, yet without gains in data usefulness.\fi Overall, our findings show that coarsening data directly erodes its value, and highlight the need for using data-coarsening, not as stand-alone mechanism, but in combination with data-sharing models that provide adjustable degrees of accountability and security.
\end{sciabstract}

\tableofcontents

\pagebreak
\section{Introduction}
%\vspace{8pt}

Large-scale datasets of human behavior are likely to revolutionize the way we develop cities, fight disease and crime, and respond to natural disasters. However, these consist of sensitive information, such as citizens' geo-location, purchasing behavior, and socialization patterns. Moreover, numerous studies have shown adversarial methods that can successfully associate sensitive information in anonymized datasets to individual identities\textemdash i.e., reidentification \cite{de2013unique,gramaglia2014anonymizability,song2014not,de2015unique, cecaj2016re, boutet2016uniqueness, narayanan2008robust, el2011systematic, wondracek2010practical}. Hence, understanding and managing the risk to privacy of these datasets still preconditions their broad use and potential impact.

In this work, we consider mobile phone metadata as a paradigmatic example of what is colloquially referred to as `big data'. Due to its high granularity, high dimensionality, passive data generation process\iffalse\footnote{As opposed to survey data, for example, where subjects are explicitly asked to provide information.}\fi, and high potential value, mobile phone metadata represents most characteristic features of novel data types at the core of `big data'. Other such types include GPS tracks, web browsing, financial behavior, genetic data, and satellite imagery; which share in common a high potential societal value and concern for individuals' privacy. 

\subsection{\textbf{Mobile Phone Data for Development}}
\vspace{5pt}

Metadata is data about data. In the context of mobile phone usage, this represents a record that a call was made---including a time stamp and geographic location, with precision determined by the location of cell towers---but no information on the content of the call itself. Mobile phone metadata is most commonly referred to as CDRs (Call Detail Records)\iffalse \footnote{Note that for a precise terminology one would distinguish CDRs from DDRs (Data Detail Records). However, most often the former is used as if it includes the latter.}\fi. Table \ref{tab:cdr} shows dummy examples of call detail records of a couple of phone calls.

\begin{table}[h]
\caption{Example of CDR records.}
\vspace{-.5cm}
\begin{center}
\begin{threeparttable}
\begin{tabular}{c| c c c c}  
    \toprule
    \textbf{ caller ID } & receiver ID & tower ID & time  \\ \midrule
    %      ZIPs   &          \%           &          \%        &     \%        \\ 
     \scriptsize{299C20B41B32B5GH76C343  } & \scriptsize{AEA595D43E2C9EE20EC12R}      &  \footnotesize{768}    &  \footnotesize{16-12-03 16:50}      \\
     \scriptsize{299C20B41B32B5GH76C343  } & \scriptsize{C721FD9F5A8902BD1EE9C4}  &   \footnotesize{981}   &  \footnotesize{16-12-24 19:56}   \\
     \scriptsize{B8673E7C673FC9EZ958FB6  } & \scriptsize{3ACC4FDDD29B45ZX1A2012}  &  \footnotesize{255}  &  \footnotesize{16-12-24 20:34}   \\ \bottomrule
\end{tabular}
%\begin{tablenotes}     % add notes with \tnote{1} 
%\item[1] qwerty; \item[2] asdfgh
%\end{tablenotes}
\end{threeparttable}
\end{center}
\label{tab:cdr}
\end{table}

Relevant characteristics of CDRs are: 1) the caller and receiver identities are pseudonymized, i.e., their phone numbers are replaced by anonymous pseudonyms (e.g., through hashing); and 2) the geographic location of towers used for each communication provide an approximation of the user's location. 

CDRs are a particularly pervasive and relevant data source for development and humanitarian response purposes. They are generated by standard telecommunication infrastructure, and collected by mobile phone companies on an ongoing basis. Moreover, handsets and airtime are becoming cheaper, leading to increased penetration and representativity, which by 2013 approached 89\% in developing countries and 96\% globally\cite{UNmobile2013}. 

There are several ways in which the location information in mobile phone metadata is analyzed and used. For example, it is possible to build dynamic maps of population density and population mobility in real time, over areas as large as countries, and at high geographic and individual detail \cite{deville2014dynamic}. This information in turn has valuable applications in a wide range of development and humanitarian action domains, such as: disaster response upon earthquakes and floods \cite{bengtsson2011improved,cdrs-tabasco}, epidemic analysis of malaria and influenza outbreaks \cite{wesolowski2012quantifying,frias2011agent}, socio-economic and poverty mapping in both the developed and developing worlds \cite{eagle2010network,blumenstock2015predicting,steele2017mapping}, transportation systems development \cite{berlingerio2013allaboard}, and improving national statistics \cite{jahani2017improving}.

%%%%%%%%%%%%%%%%%%%%%%%%%%%%%%%%%%%%%%%%%%%%%%%%%%%%%
\subsection{\textbf{Privacy Risk in Mobile Phone Metadata}}
\vspace{5pt}

\iffalse Psuedonoymized metadata, and CDRs in particular, is metadata in which all personal information such as names, addresses and telephone numbers are replaced with unique and anonymous identifiers (see caller IDs in Table~\ref{tab:cdr}). This means that the sequence of records of an individual user continue to be associated with one another (e.g., first two records in Table \ref{tab:cdr}), but no explicit connection can be made between the sequence of records and the actual identity of the individual.\fi

In the recent past, privacy provided by pseudonymization coupled with institutional non-disclosure agreements (NDAs) has served as basis to allow sharing of large CDR datasets. However, research has recently shown adversarial methods that successfully associate sensitive information in the datasets to individuals' identities, even under pseudonymization of all personal identifiers \cite{de2013unique, gramaglia2014anonymizability,song2014not, de2015unique, cecaj2016re, boutet2016uniqueness, narayanan2008robust,el2011systematic,wondracek2010practical}.

A seminal study on reidentification of CDRs analyzed mobility data from 1.5 million mobile phone subscribers in a small western country, where the location of an individual was specified hourly and with a spatial resolution given by the geographic distribution of the carrier's antennae \cite{de2013unique}. It demonstrated that outside knowledge of just four random spatio-temporal points were enough to uniquely identify 95\% of individuals in the database. Furthermore, the study showed that data can be coarsened in order to reduce the likelihood of re-identification. This coarsening, more properly named \textit{spatial and temporal generalization}, is a key technique applied to data to preserve privacy, allowing companies, NGOs, and public organizations to balance privacy risks with data's potential for positive societal impact.
\iffalse Data transformations of this kind, coupled with data sharing models of varying levels of security \textemdash such as open algorithm platforms, limited remote access, question and answer APIs, and legal non-disclosure agreements (NDAs) \cite{de2014d4d}\textemdash constitute the fundamental control variables that allow policy to balance and tradeoff between data privacy and its potential for positive societal impact.\fi

%%%%%%%%%%%%%%%%%%%%%%%%%%%%%%%%%%%%%%%%%%%%%%%%%%%%%%%%%%%%%%%%%%%%%%%%%%%%%%%%%
%%%%%%%%%%%%%%%%%%%%%%%%%%%%%%%%%%%%%%%%%%%%%%%%%%%%%%%%%%%%%%%%%%%%%%%%%%%%%%%%%%%%%%%%
\section{Methods}
\vspace{10pt}

%%%%%%%%%%%%%%%%%%%%%%%%%%%%%%%%%%%%%%%%%%%%%%%%%%%%%
\subsection{\textbf{Assessing Privacy}}
\vspace{4pt}

%%%%%%%%%%%%%%%%%%%%%%%%%%%
\vspace{.15cm}
\noindent\textbf{\normalsize{Concepts and Vocabulary}}
\vspace{.2cm}

%\comment{this section can be shrinked/removed depending where we submit}

\noindent Datasets contain attributes such as name, telephone, address, income, health status, location, items purchased, and websites visited. These attributes can be classified as: \textit{direct identifiers}, \textit{quasi-identifiers}, or \textit{sensitive attributes}. For example, in an anonymous health database, where names and social security numbers (direct identifiers) are pseudonymized, a prying third party with access to the database could attempt to know the medical condition (sensitive attributes) of Jane by using auxiliary information of her ZIP code and age (quasi-identifiers) to single her out. We denote the set of auxiliary information about quasi-identifiers of person $i$ by $a_i$; and refer to the subset of individuals whose records match $a_i$ as the \textit{equivalent class} of $i$ given $a_i$, denoted by $E_i$. Jane is reidentified if $|E_i|=1$. \iffalse(see example in Appendix I).\fi

Traditional measures to protect privacy have focused on guaranteeing that an attacker, even with full knowledge of an individual's quasi-identifiers, is unable to reidentify her uniquely\cite{priv-kanonymity02}, or extract information about her\cite{machanavajjhala2006ell,li2007t}. For example, a privacy approach of widespread use in the last decade is \textit{k-anonymity} \cite{priv-kanonymity02}, where the granularity of quasi-identifiers is gradually reduced, thus increasing the size of equivalent classes, until the requirement $\min_{\forall i} |E_i| \geq k$ is met\iffalse(see example in Appendix I)\fi. These approaches, however, are incapable of coping with most behavioral datasets due to their high-dimensionality\cite{noriega2015balancing}, raising the need for appropriate ones.

%%%%%%%%%%%%%%%%%%%%%%%%%%%
\vspace{.3cm}
\noindent\textbf{\normalsize{Privacy in High-Dimensional Data}}
\vspace{.2cm}

\noindent People's online activity leaves a comprehensive data trace behind, which coupled with the advent of technologies for pervasive sensing, amount to an unprecedented instrumentation of our societal systems. Notably, data at the core of today's  ``big data" is high-dimensional. Examples are human mobility data, banking and credit card data, consumer behavior, web browsing, online social networks, and genetic data.

High-dimensional datasets contain only a sparse sample of the space of possible records, which, similar to fingerprints, often entails that individual records are unique. To illustrate how sparsity can be exploited for reidentification, consider a very large database of song lyrics. The space of all possible song lyrics---permutations of a bounded number of words---is extremely large. Thus, given a sequence of only 3 or 4 words, we are likely to identify a song uniquely among thousands. In practice, research has shown high reidentifiability in varied high-dimensional datasets, from mobile phone records and credit card transactions to online movie reviews \cite{de2013unique,de2015unique,narayanan2008robust}.

%%%%%%%%%%%%%%%%%%%%%%%%%%%
\vspace{.3cm}
\noindent\textbf{\normalsize{Measures of Reidentifiability}}
\vspace{.2cm}

\iffalse High dimensionality renders \textit{k-anonymity} and similar approaches inappropriate for reidentification analysis \cite{noriega2015balancing}.\fi

\noindent Measures for assessing reidentification risk in high-dimensional data must recognize sensitive attributes themselves as quasi-identifiers, and vice-versa, moving to a paradigm of partial adversary knowledge as basis of reidentification. One such measure is \textit{unicity} \cite{de2013unique}. The unicity $u_p$ of database $D$ is calculated as the percentage of users in $D$ who are reidentified by using $p$ randomly selected data points from each user's records; i.e., the percentage of users whose equivalent class satisfies $|E_i^p|=1$. For instance, it was shown that outside knowledge of four calls was enough to reidentify 95\% out of 1.5 million individuals in a CDR dataset ($u_{4}=95\%$).

Here we elaborate on previous work and propose the following two metrics of privacy in high-dimensional datasets. We aim at metrics that are meaningful and intuitive, as well as rooted in the formal framework of information theory.

\vspace{.2cm}
\textbf{\textit{Information cost}}. Similar in spirit to \textit{unicity}, we define the \textit{information cost} of reidentification in $D$ as the average quantity of outside information that suffices to reidentify users in $D$. Let $c_i$ denote the number of data points drawn from user $i$'s records needed to reidentify her, then information cost of D is defined as $c = \frac{1}{n} \sum c_i$, where $n$ is the number of users.

\vspace{.2cm}
\textbf{\textit{Information ratio}}. In addition, we define the \textit{information ratio} $r$ of $D$ as the average fraction of a user's data that is required to reidentify her. Let $|d_i|$ with $d_i \in D$ denote the amount of $i$'s data in $D$, then the \textit{information ratio} of $D$ is given by $r = \frac{1}{n} \sum \frac{c_i}{|d_i|}$. Relevantly, the information ratio summarizes not only the amount of information needed for reidentification, but also the amount of information gained by an adversary once a user is reidentified; where\hspace{3pt}  $1 - r$  \hspace{1pt} is the average \textit{information gain}. This feature of the \textit{information ratio} is highly relevant, as it enables stakeholders to reflect over preferences accounting for both key elements of privacy risk: information requirement and information gain.     

\vspace{.2cm}
These measures connect with information theory through the core concept of average information content\textemdash i.e., the \textit{entropy} of multivariate distributions \cite{mackay2003information}. In particular, the higher the entropy of a dataset, the higher the information content of any bit of adversary knowledge, and hence the fewer bits of information required for reidentification (lower \textit{information cost} and \textit{ratio}). Moreover, the measures convey a meaningful and intuitive interpretation, which may help a broader audience reflect upon and assess both the likelihood and potential harm that reidentification entails. Below we apply these measures to the case of CDRs at several spatio-temporal granularity levels.

\begin{figure}[h]
\centering
\includegraphics[width= .8\columnwidth]{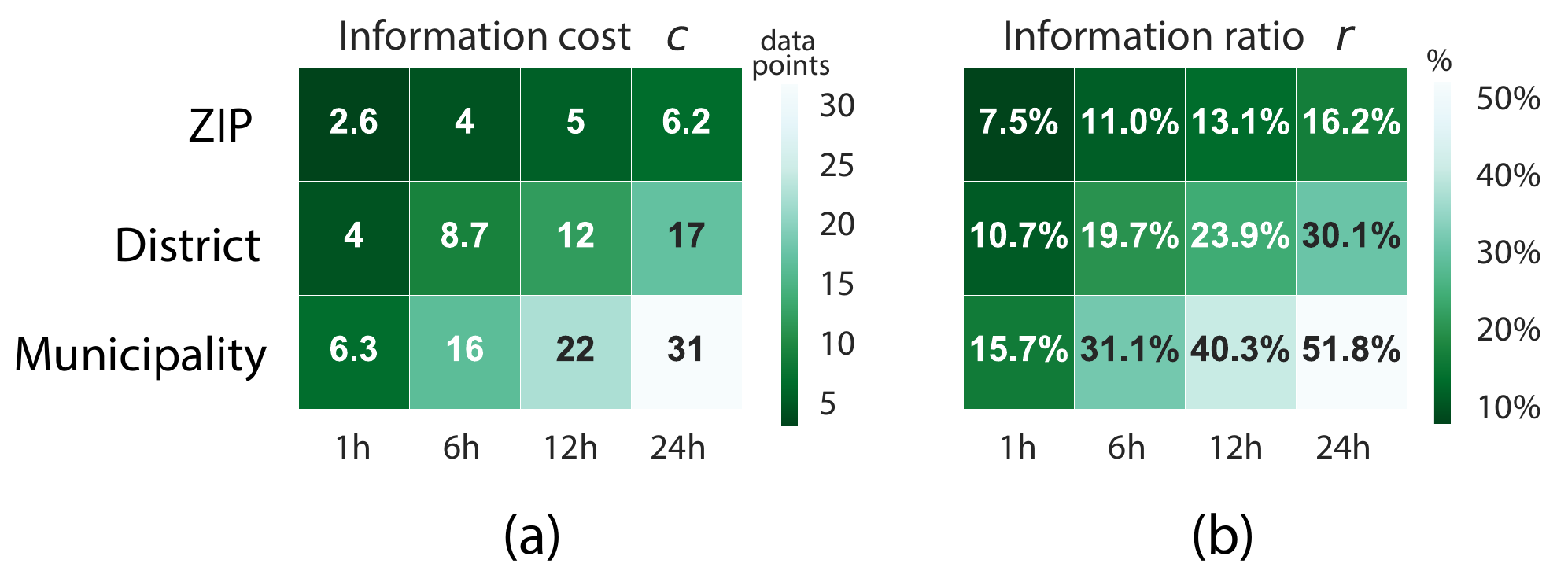}
\caption{ \textbf{Reidentification results.} Reidentification results for datasets at varying levels of spatial and temporal granularity: (a) \textit{information cost} and (b) \textit{information ratio} . All 95\% confidence intervals are non-overlapping (except for pairs ($D1$, $Z6$) and ($D24$, $M6$), as shown in Additional File 1).}
\label{fig:privacy}
\end{figure}

%%%%%%%%%%%%%%%%%%%%%%%%%%%%%%%%%%%%%%%%%%%%%%%%%%%%%
\subsection{\textbf{Assessing Utility}}
\label{sec:methods-usefulness}
\vspace{8pt}

In order to assess the usefulness of mobile phone data at the various spatial and temporal generalization levels, we collected data from a  quantitative survey targeted to experts with experience in research and analysis of mobile phone data for development and humanitarian action. In  particular, the survey's population were experts who took part in D4D-Senegal 2014, the open innovation data challenge based on anonymous records of Orange's mobile phone users in Senegal \cite{de2014d4d}. Thirty two D4D experts\textemdash members of academia and research institutions around the globe\textemdash opted in to respond the survey. Notably, the pool represented a diversity of twenty five research institutions, across fourteen countries and five continents; and a spread of domain foci in health, transportation and urban planning, national statistics, and others (see Table~\ref{tab:survey-stats}).

\begin{table}[h]
\caption{Experts data.}
%\vspace{-.3cm}
\begin{center}
\begin{tabular}{@{}|c|c|@{}}
\midrule
      \textbf{Number of experts}    &   32    \\   \midrule
      \textbf{Number of institutions}   &   25  \\     \midrule
      \textbf{Continents}   &          North America, South America, Asia, \\ &  Africa and Europe.         \\      \midrule
      \textbf{Countries}   &           Belgium, Cameroon, Canada, Chile, China\\ &  France, Germany, India, Italy, Japan, \\ & Spain, Sweden, United Kingdom, USA. \\ \midrule
      \textbf{Domain focus of respondents}   &         Health 20.5\%, Transportation and Urban \\ & Planning 34\%,  National Statistics 20.5\%, \\ & and Others 25\%
\\ \bottomrule
\end{tabular}
\end{center}
\label{tab:survey-stats}
\end{table}

The survey asked experts to consider a scenario in which they were provided with CDRs from a large metropolitan region in the developing world, including all call communications of a large representative sample of the population in the region. Experts rated on a scale from 1 to 10 the usefulness of such data in their research domains, if provided generalized at the various spatio-temporal granularity levels shown in Figure~\ref{fig:privacy} (screenshots in Additional File 2). \iffalse Participation in the survey was voluntary and kept completely anonymous (survey screenshots are found in Additional File 2).\fi

%%%%%%%%%%%%%%%%%%%%%%%%%%%%%%%%%%%%%%%%%%%%%%%%%%%%%%%%%%%%%%%%%%%%%%%%%%%%%%%%%
\section{Results}
\vspace{8pt}

%%%%%%%%%%%%%%%%%%%%%%%%%%%%%%%%%%%%%%%%%%%%%%%%%%%%%
\subsection{\textbf{Reidentification Results}}
%\label{sec:results_priv}
\vspace{8pt}

We analyzed a mobile phone dataset $D$ comprising phone calls of 1.4M people across a large metropolitan region in the developing world over a month in 2013. From it we derived generalized datasets for each combination of spatial and temporal granularity levels $g \in \{ ZIP,$ $District, Municipality\} \times \{ 1h, 6h, 12h, 24h \}$.

The spatial granularity levels used were ZIP code, district, and municipality  polygons, which partition the space in 56, 156, and 2130 polygons with average areas of 101, 36, and 3 km$^2$ respectively. The temporal granularity levels used were time slices with duration of 1 hour, 6 hours, 12 hours, and 24 hours. For example, under dataset $D_{Z6}$ ---generalized at ZIP code and 6 hours granularity--- a call issued at 4 pm from ZIP code 02139 by one user is indistinguishable from a call issued at 7 pm by another user in the same ZIP code. We computed the \textit{information cost} $c(D_g)$ and \textit{information ratio} $r(D_g)$ of reidentification associated to each generalized dataset $D_g$. Figure \ref{fig:privacy} shows the results. 

% figure fig:privacy

We observe that for the most granular dataset, $D_{Z1}$, it takes on average 2.6 bits\textemdash i.e., data points\textemdash to reidentify an individual, which represents 7\% of that individual's data \big( $c(D_{Z1}) = 2.6$ and $r(D_{Z1} ) = 7\%$ \big). This means that a prying third party with outside knowledge of 7\% of an individual's data could reidentify her and obtain the remaining 93\%. In contrast, we observe that for the least granular dataset $D_{M24}$, reidentification requires an average of 32 data points, or 51\% of an individual's data \big( $c(D_{M24}) = 32$ and $r(D_{M24} ) = 51\%$\big).
\iffalse \footnote{Consistent with previous studies \cite{de2013unique,noriega2015balancing}, unicity at four points increased with data granularity, and ranged $u_4(D_g) \in [97\%,1.5\%]$ (see \nameref{S1_Fig})} \fi
Hence, if $D_{M24}$ is published, a prying third party requires on average outside knowledge of about 51\% of an individual's data to reidentify her and obtain the remaining 49\%.

%%%%%%%%%%%%%%%%%%%%%%%%%%%%%%%%%%%%%%%%%%%%%%%%%%%%%
\subsection{\textbf{Utility Results}}
\label{sec:results_use}
\vspace{8pt}

\begin{figure}[h!]
\centering
\includegraphics[width= .8\columnwidth]{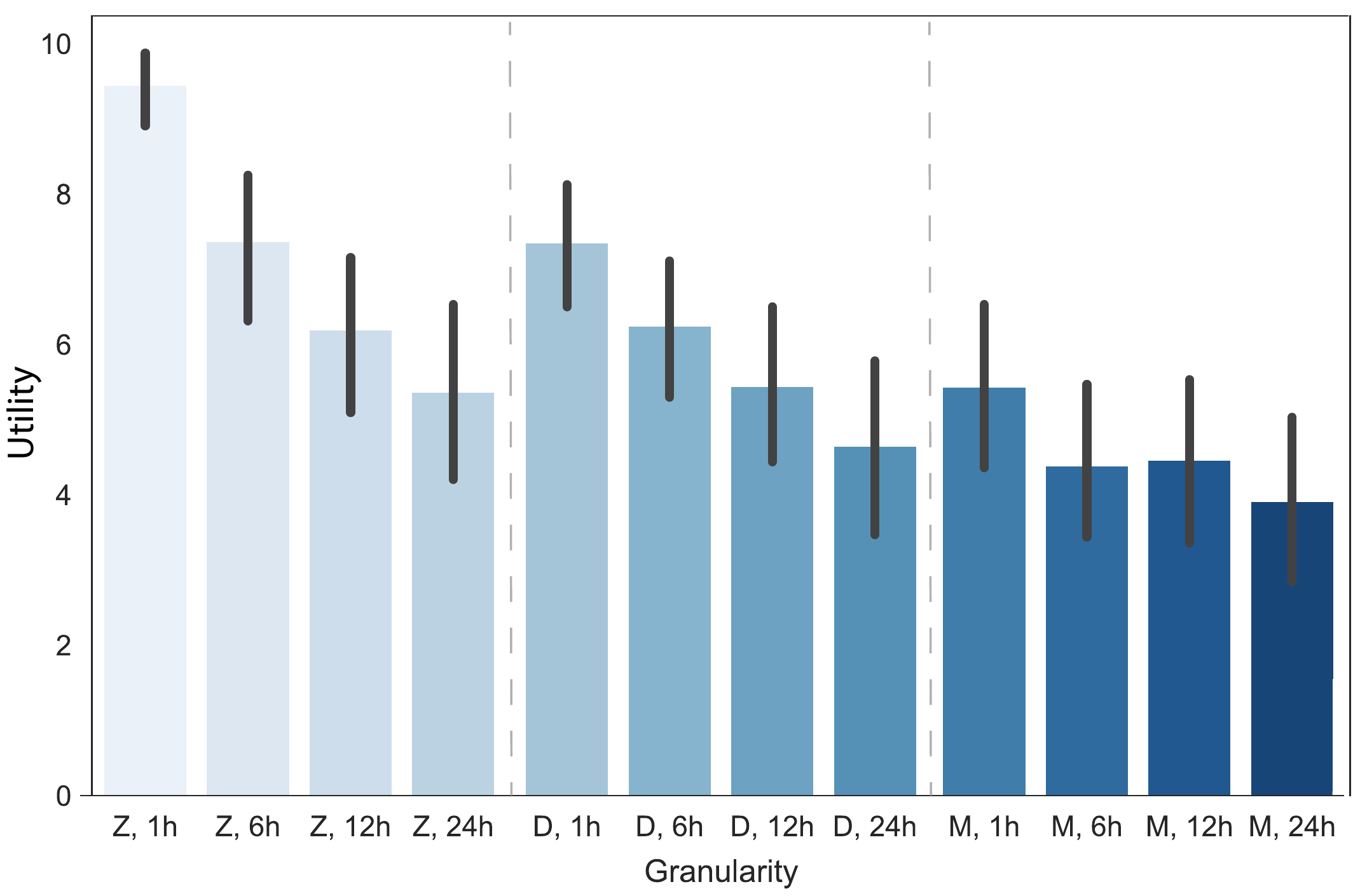}
\caption{\textbf{Utility results.} Usefulness of mobile phone metadata per generalization profile. Grey bars denote bootstrapped 95\% confidence intervals. Spatial granularity levels are Z = ZIP code, D = district, and M = municipality. }
\label{fig:usefulness}
\end{figure}

Figure \ref{fig:usefulness} shows the experts’ assessment of data utility for each granularity level. We observe that data usefulness decays as the data is generalized spatially and temporally, with values ranging from 9.3 to 4.0 for the most and least granular datasets ($D_{Z1}$ and $D_{M24}$).

%%%%%%%%%%%%%%%%%%%%%%%%%%%%%%%%%%%%%%%%%%%%%%%%%%%%%
\subsection{\textbf{Privacy-Utility Tradeoff}}
%\label{sec:results_basic}
\vspace{8pt}

\begin{figure}[h!]
\centering
\includegraphics[width= .85\columnwidth]{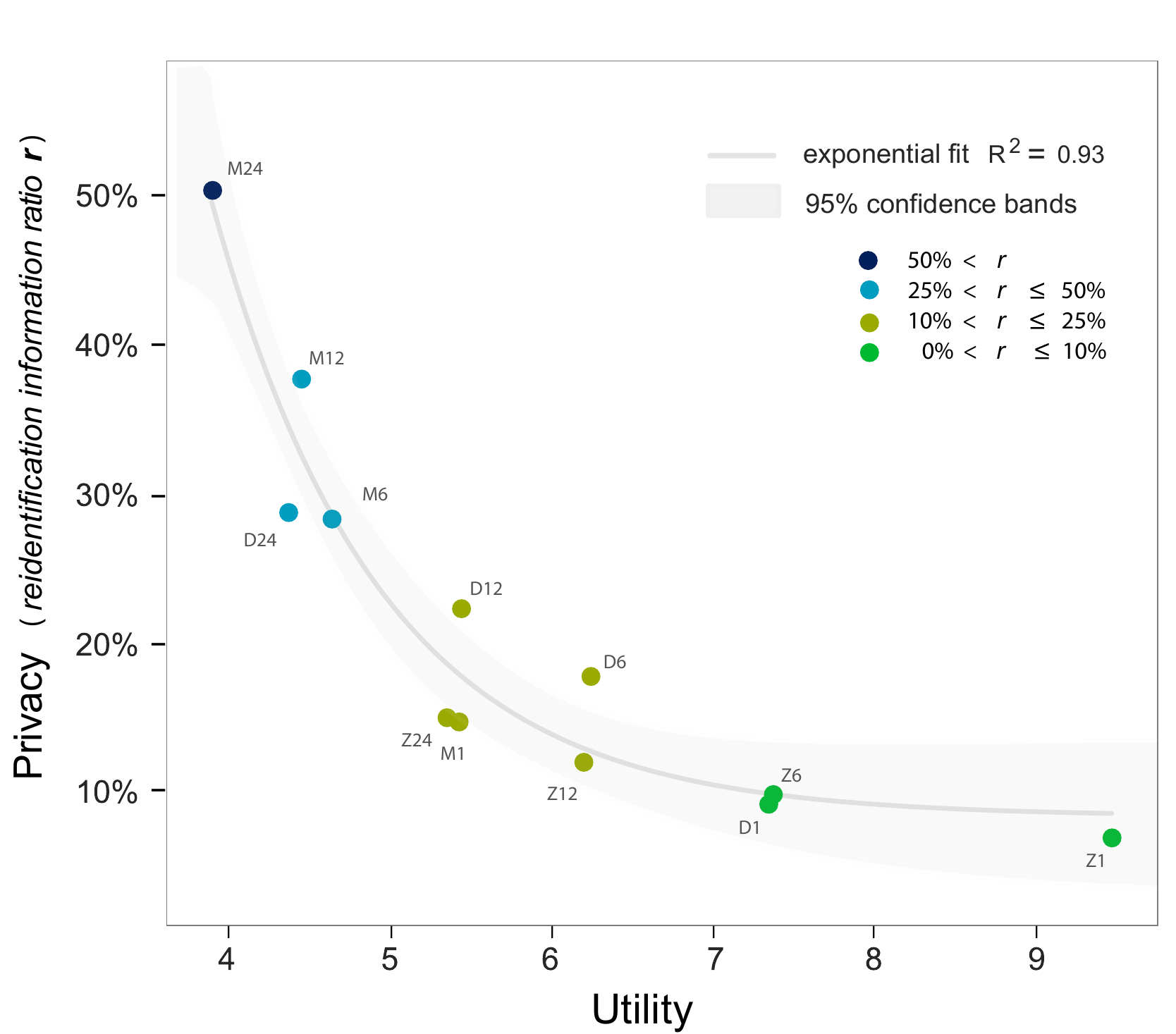}
\caption{\textbf{Privacy-Utility trade-off in mobile phone data.} Utility vs. reidentification risk in mobile phone data for development, across spatial and temporal granularities $\{ ZIP, District, Municipality\}\times\{ 1h, 6h, 12h, 24h\}$. The more useful the dataset, the less auxiliary information is needed to reidentify its individuals. Conversely, while data  generalization increasingly hinders reidentification, it strongly diminishes datasets' value.}
\label{fig:tradeoff}
\end{figure}

Figure \ref{fig:tradeoff} shows results on the privacy-utility tradeoff. Each point represents a generalized dataset $D_g$ assessed on usefulness and reidentification risk, where the optimal position corresponds to the top right corner\textemdash high usefulness and hard reidentification.

We observe a sharp tradeoff between usefulness and privacy. The most granular dataset $D_{Z1}$ is the most valuable, with usefulness score of 9.3; however, it is also the dataset most prone to reidentification, where on average a third party with outside knowledge of only 7\% of an individual's data can reidentify the individual and gain the remaining 93\% of personal information. Conversely, the least granular dataset $D_{M24}$ is the least valuable, with usefulness score of 4; yet it's also the dataset least prone to reidentification, where on average a third party requires outside knowledge of 51\% of an individual's data to gain the remaining 49\% of personal information.

Figure \ref{fig:tradeoff} also shows that the tradeoff is not strict. Generalization levels such as $D24, Z24, M1,$ and $Z12$ are Pareto-suboptimal, or dominated. For example, $D24$ and $M24$ have similar usefulness, however an adversary requires about 65\% more outside information to reidentify an individual in $M24$ than in $D24$ \big( $r(M_{24}) = 51\%$ and $r(D_{24}) = 29\%$  \big). 

The tradeoff in Figure \ref{fig:usefulness} implies that, while generalization increasingly hinders reidentification, it strongly undermines data utility. This highlights the complimentary roles of coarsening and data sharing models in enabling use while controlling risks. For example, datasets most prone to reidentification---such as $D1, Z6,$ and $Z1$, with reidentification information ratio $r \leq 10\%$---could be shared only under strict models, such as precomputed indicators, or the use of open algorithm platforms \cite{hardjono2016trust,opalwebsite}. Figure \ref{fig:usefulness} also implies that even highly coarse datasets can be vulnerable to reidentification, and hence should not be made fully public. However, we may want to share more broadly datasets posing more moderate reidentification risks---such as $M24$, with reidentification information ratio $r > 50\%$---, through models similar to those used in D4D challenges \cite{de2014d4d}, where data is accessed by a limited number of semi-trusted parties under non-disclosure agreements (NDAs). Similarly, datasets posing moderate-high reidentification risks---such as those with $10\% < r < 50\%$---, could be shared under additional control mechanisms, such as remote access with adjustable disclosure controls via Q\&A architectures, and/or accountability and deterring incentive schemes \cite{wan2015game}. See \cite{d2015privacy,yves17} for details and discussion on modern data sharing models and protocols.

%%%%%%%%%%%%%%%%%%%%%%%%%%%%%%%%%%%%%%%%%%%%%%%%%%%%%%%%%%%%%%%%%%%%%%%%%%%%%%%%%
\section{Conclusions}
\vspace{8pt}

The present work shows for the first time the notorious trade-off between the societal value of mobile phone data for development and humanitarian action, and the reidentification risk to which individuals in it are exposed. Because data generalization directly erodes data's value, it cannot be regarded as a silver bullet solution for preserving privacy in high-dimensional datasets \cite{narayanan2014no}. Yet, coupled with data-sharing models that provide adjustable degrees of accountability and security, it may help find the right balance between privacy and utility.

This work assessed data utility as the value provided to experts in the analysis of mobile phone data for development and humanitarian action. This approach is particularly germane when considering purpose-specific data sharing, such as in the case of poverty mapping, transportation planning, or assisting response efforts upon natural disasters. We anticipate future work focusing on the trade-offs of mobile phone data usage in alternative domains, such as marketing and credit scoring.

The formal measures of reidentification risk here proposed can provide meaningful and intuitive summaries of the information requirements, and information gains, associated with reidentification. \iffalse We anticipate information-theoretic extensions of this work, such as development of metrics that capture information continuity along spatiotemporal and social network spaces.\fi
Ultimately, we hope this work helps promote participation of broader audiences in reflecting upon data privacy tensions, as societal preferences are indispensable inputs for resolving where systems should sit along the privacy-utility spectrum.

\iffalse
(Limitations:) This study focuses on mobile phone metadata, hence, although previous results have been shown valid across domains (e.g., across CDRs and credit card records \cite{de2013unique,de2015unique})   further studies are required to extend these conclusions to domains such as credit card, health, and web browsing records.
\fi

%%%%%%%%%%%%%%%%%%%%%%%%%%%%%%%%%%%%
% Bibliography %%%%%%%%%%%%%%%%%%%%%%%%%%%%%%%%%%%% %%%%%%%%%%%%%%%%%%%%%%%%%%%%%%%%%%%% %%%%%%%%%%%%%%%%%%%%%%%%%%%%%%%%%%%% %%%%%%%%%%%%%%%%%%%%%%%%%%%%%%%%%%%%
%\bibliographystyle{Science}
\bibliographystyle{ieeetr}
\bibliography{references}

\vspace{1cm}
%\pagebreak
\noindent\textbf{Acknowledgements}

\noindent The authors are grateful to Robert Kirkpatrick, Mila Romanoff, and Miguel Luengo-Oroz for fruitful discussions regarding data privacy throughout the years. We also thank Julie Ricard for her valuable input in preparing this manuscript. \\

\noindent\textbf{Funding}

\noindent This work was supported by funds from the MIT Trust Data Consortium (trust.mit.edu). All findings and conclusions are those of the authors and do not necessarily reflect the views of their sponsors, institutions, and colleagues. Alejandro Noriega-Campero was partially supported by the Mexican Council for Science and Technology (CONACYT). \\

\noindent\textbf{Competing interests}

\noindent The authors declare that they have no competing interests. \\

\noindent\textbf{Author contributions}

\noindent ANC, YAM, AR, and AP designed the research. ANC and OL performed the analysis and obtained the results. ANC and AP wrote the paper.  \\

\noindent\textbf{Availability of data and materials}

\noindent The datasets necessary to reproduce all figures supporting the conclusions of this article are included in Additional files 3 and 4 available upon request. All data analyzed was anonymous, removed from all location data, and only aggregate reidentification statistics were computed. \\

\noindent\textbf{List of abbreviations}

\noindent GPS, global positioning system; CDR, call detail records.

%%%%%%%%%%%%%%%%%%%%%%%%%%%%%%%%%%%%%%%%%%%%%%%%

\includepdf[pages=-]{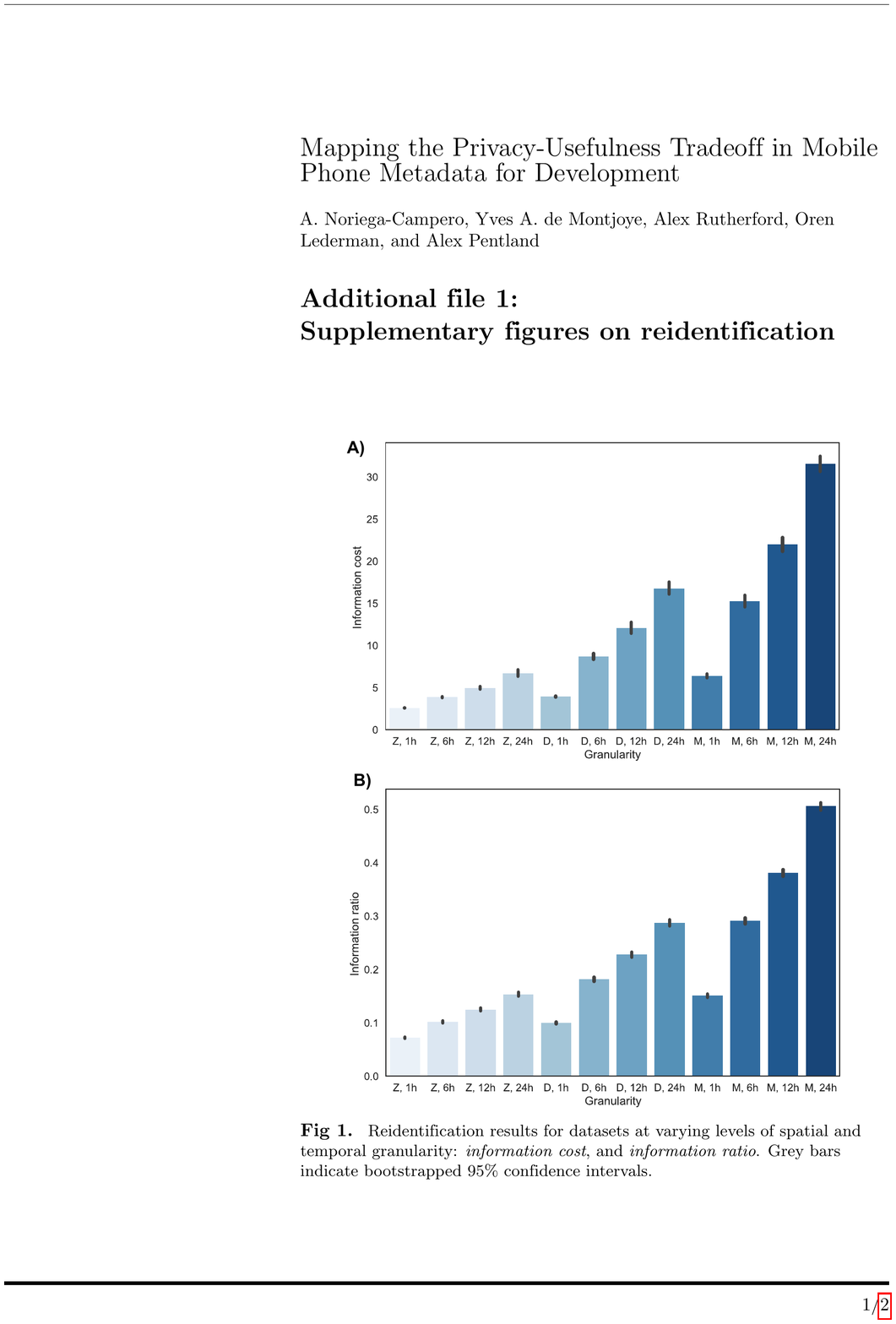}

\end{document}